\documentstyle[11pt,appb,epsfig]{article}        %
\begin{document}
\title{EQUATION OF STATE OF NUCLEONIC MATTER\footnote{Supported by BMBF and GSI
Darmstadt}}
\author{W. Cassing\thanks{In collaboration with A. Hombach, U. Mosel and P. K. Sahu} \\ Institut f\"ur
Theoretische Physik, Universit\"at Giessen \\ D-35392 Giessen,
Germany}
\date{ }
\maketitle

\newcommand{\be}{\begin{equation}}
\newcommand{\ee}{\end{equation}}
\newcommand{\bea}{\begin{eqnarray}}
\newcommand{\eea}{\end{eqnarray}}
\newcommand{\aein}{\hspace*{.5cm}}
\renewcommand{\d}{\partial}
\newcommand{\nn}{\nonumber\\}
\newcommand{\BUU}{Boltzmann-Uehling-Uhlenbeck}

\begin{abstract}

The nuclear equation of state (EoS) is investigated by flow
phenomena in relativistic heavy-ion collisions, both in transverse
and radial direction, in comparison to experimental data from 150
A MeV to 11 A GeV. To this aim the collective dynamics of the
nucleus-nucleus collision is described within a transport model of
the coupled channel RBUU type. There are two factors which
dominantly determine the baryon flow at these energies: the
momentum dependence of the scalar ($U_S$) and vector potentials
($U_\mu$) for baryons and the resonance/string degrees of freedom
for energetic hadron excitations. We fix the explicit momentum
dependence of the nucleon-meson couplings by the nucleon optical
potential up to 1 GeV and extrapolate to higher energy. When
assuming the optical potential to vanish identically for $E_{kin}
\geq 3.5$ GeV we simultaneously describe the sideward flow data of
the PLASTIC BALL, FOPI, EoS and E877 collaborations, the elliptic
flow data of the E895 and E877 collaborations and approximately
the rapidity and transverse mass distribution of protons at AGS
energies without employing any {\it explicit} assumption on a
phase transition in the EoS. However, the gradual change from
hadronic to string degrees of freedom with increasing bombarding
energy can be viewed as a transition from {\it hadronic} to {\it
string} matter, i.e. a dissolution of hadrons at high energy
density.
\end{abstract}


\section{Introduction}
Relativistic heavy-ion collisions (RHIC) provide a unique tool to
study nuclear matter at high densities and temperatures,
reminiscent of the early big-bang of the universe, but with better
statistics and under controlled conditions. These reactions also
provide constraints on the interior of neutron stars, where the
nuclear equation-of-state (EoS) plays an essential role for the
possible existence of an inner quark core or an extended mixed
phase of quarks and hadrons \cite{Schertler}. However, since in a
RHIC the system initially is far away from thermal and chemical
equilibrium, both particle production and collective motion depend
on various quantities such as the stiffness of the EoS, the
momentum dependence of the interaction or mean-field potentials
(MDI), in-medium modifications of the $NN$ cross section
$\sigma_{NN}$, the initial momentum distribution of the nucleons
\cite{GBdG87,BKL91} as well as the number of hadronic degress of
freedom accounted for in the transport simulation \cite{hom98}. It
is thus necessary not to focus on a single observable alone but to
investigate the dynamical evolution of the RHIC within a single
model that is able to describe all relevant single-particle as
well as collective quantities.

Whereas the experimental and theoretical studies of collective
nuclear flow have been restricted to the 1-2 A GeV energy regime
in the past \cite{Stoecker,gut89}, more recently both the directed
transverse flow (sideward flow) and the flow tensor (elliptic
flow) have been measured and reported by the BNL-E877
collaboration \cite{her96,rei97,E877,E895} for heavy-ion ($Au+Au$)
collisions at AGS energies in the energy range of 1 A GeV $\le
E_{inc} \le 11 A$ GeV. In this energy range the directed
transverse flow first grows, saturates at around 2 A GeV, and then
decreases experimentally with energy showing no minimum as
expected from hydrodynamical calculations including a first order
phase transition in the EoS \cite{Rischke}. Whether this decrease
in directed flow or the change of sign in elliptic flow is
indicative of a phase transition \cite{dani98} is a question of
high current interest.

In this contribution the collective behaviour of nuclear matter in
a heavy-ion collision is reviewed in the energy range from 150 A
MeV to 11 A GeV for various systems using the transport model
\cite{sahu98}. For energies above 1 A GeV it has been,
furthermore, complemented by the string dynamics from the HSD
transport approach \cite{ehe96} which has been tested extensively
for $p+A$ and $A+A$ collisions from SIS to SPS energies
\cite{cass99}.

\section{The extended RBUU-Model}
To describe the heavy-ion collision data at energies starting from
the SIS at GSI to the SPS regime at CERN, relativistic transport
models have been extensively used
\cite{ehe96,Dani,mar94,bas97,li97}. For a general derivation of
transport theories the reader is refered to Ref. \cite{Cass} and
to Ref. \cite{cass99} for a recent review. Among these transport
models the Relativistic Boltzmann-Uehling-Uhlenbeck (RBUU)
approach incorporates the relativistic mean-field (RMF) theory,
which is applicable also to various nuclear structure problems as
well as for neutron star studies \cite{Schertler,SahuNS}. Since it
is based on an effective hadronic Lagrangian density it allows to
evaluate directly the nuclear EoS at zero and finite nuclear
temperature $T$ as well as the scalar and vector mean fields $U_S$
and $U_\mu$, that determine the in-medium particle properties, for
arbitrary configurations of nucleons in phase space \cite{Cass}.
Here we essentially base the studies on the Lagrangian (parameter
set NL3) from Ref. \cite{lan91} since this Lagrangian has been
applied widely in the analysis of heavy-ion collisions by various
groups \cite{sahu98,bli99,soff99}.

We recall that the most simple versions of RMF theories assume the
scalar and vector fields to be represented by point-like
meson-baryon couplings. These couplings lead to a linearly growing
Schr{\"o}dinger-equivalent potential in nuclear matter as a
function of the kinetic energy $E_{kin}$, which naturally explains
the energy dependence of the nucleon optical potential at low
energies ($\leq 200$ MeV). However, a simple RMF does not describe
the nucleon optical potential at higher energies, where it
deviates substantially from a linear function and saturates at
$E_{kin} \approx$ 1 GeV \cite{ham90}. Since the energy dependence
of sideward flow is controlled in part by the nucleon optical
potential, the simple RMF cannot be applied to high-energy
heavy-ion collisions. In order to remedy this aspect, the more
sophisticated RBUU approaches invoke an explicit momentum
dependence of the coupling constant, i.e. a form factor for the
meson-baryon couplings \cite{ehe96,cass99,web93}.

Further important ingredients at AGS energies are the
resonance/string degrees of freedom which are excited during the
reaction in high energy baryon-baryon or meson-baryon collisions.
While at SIS energies particle production mainly occurs through
baryon resonance production and their decay, the string
phenomenology is found to work well at SPS energies ($\approx$ 200
A GeV) \cite{cass99}. One of the characteristic features of the
AGS energy regime is the competition between these two particle
production mechanisms which might be separated by some energy
scale $\sqrt{s_{sw}}$ \cite{sahu99}. Due to this complexity there
are various ways to implement elementary cross sections in
transport models~\cite{hom98,dani98,cass99,bli99,soff99,nara99}.
One of the extremes is to parameterize all possible cross sections
directly for multi-pion production, $NN \to NN n\pi \ (n \ge 3)$
only through $N, \Delta, \pi$ degrees of freedom; the other
extreme is to fully apply string phenomenology in this energy
region without employing any resonances. Although it is possible
to reproduce the elementary cross sections  from $NN$ and $\pi N$
collisions and the inclusive final hadron spectra in heavy-ion
collisions within these different models, we expect that
differences should appear in the dynamical evolution of the
system, e.g. in the thermodynamical properties \cite{hom98,nara97}
and in collective flow~\cite{sahu99}. For example, if thermal
equilibrium is achieved at a given energy density, models with a
larger number of degrees of freedom including strings will give
smaller temperature and pressure \cite{Frank1}.

We have employed here the combination of a resonance production
model \cite{effe} and the Lund string model \cite{lund} as
incorporated in the Hadron-String-Dynamics (HSD) approach
\cite{ehe96,cass99}. In the practical implementation for $NN$ or
$MN$ collisions at invariant energies lower (higher) than a
threshold energy  $\sqrt{s_{sw}}$ resonances (strings) are assumed
to be excited (see below).

\subsection{The optical potential}
In this presentation the scalar and vector mean fields $U_S$ and
$U_\mu$ are calculated on the basis of the same Lagrangian density
as considered in Ref. \cite{sahu98}, which contains nucleon,
$\sigma$ and $\omega$ meson fields and nonlinear self-interactions
of the scalar field (cf. NL3 parameter set~\cite{lan91}). The
scalar and vector form factors at the vertices are taken into
account in the form \cite{ehe96}
\begin{equation}
\label{form}
    f_s(p)=\frac{\Lambda_s^2-\frac{1}{2}p^2}{\Lambda_s^2+p^2}
    \qquad\mbox{and}\qquad
    f_v(p)=\frac{\Lambda_v^2-\frac{1}{6}p^2}{\Lambda_v^2+p^2}\ ,
\end{equation}
where the cut-off parameters $\Lambda_s = 1.0$ GeV and $\Lambda_v
= 0.9$ GeV are obtained by fitting the Schr\"odinger equivalent
potential,
\begin{equation}
\label{pot} U_{sep} (E_{kin}) = U_s + U_0 + \frac{1}{2M}
(U_s^2-U_0^2) + \frac{U_0}{M} E_{kin},
\end{equation}
to Dirac phenomenology for intermediate energy proton-nucleus
scattering \cite{ham90}.

\hskip 0.75in \psfig{figure=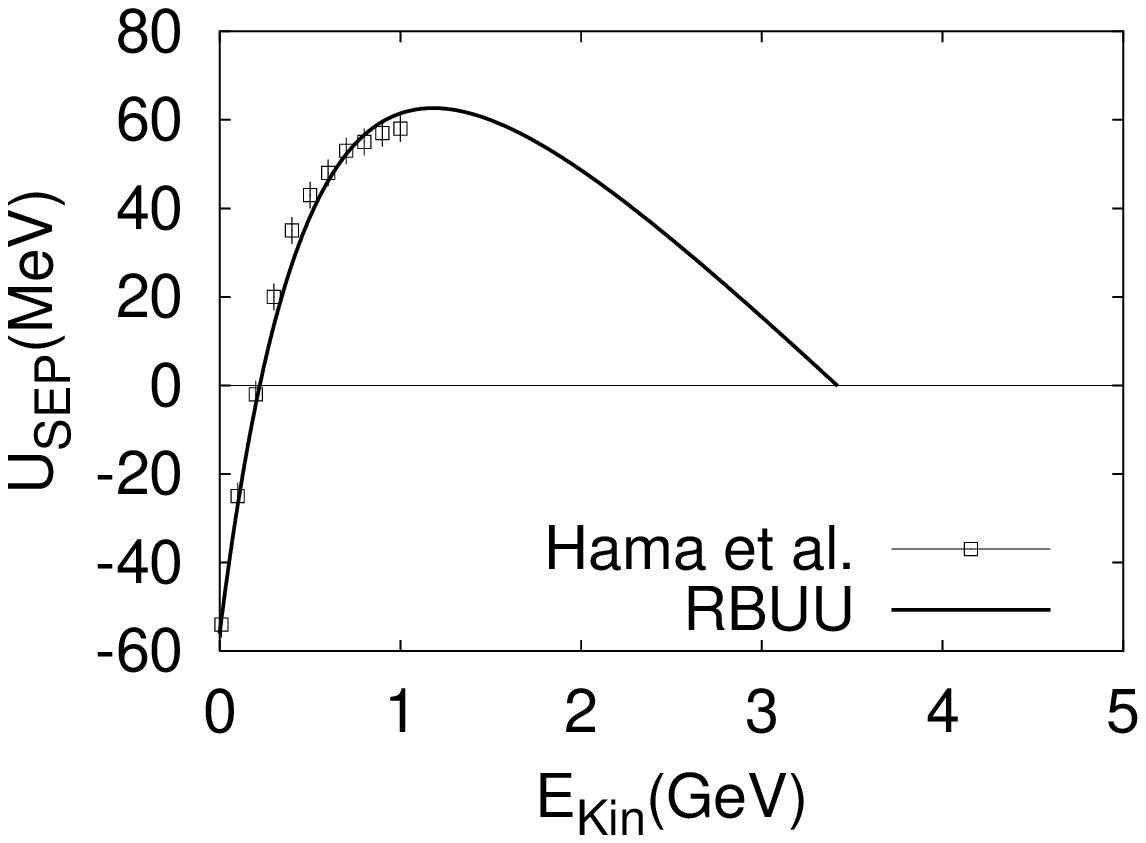,width=7cm} \vskip 0.15in
{\noindent \small {{\bf Fig. 1} The Schr\"odinger equivalent
potential (\protect\ref{pot}) at density $\rho_0$ as a function of
the nucleon kinetic energy $E_{kin}$. The solid curve (RBUU)
results from the momentum-dependent potentials discussed in the
text. The data points are from Hama et al. \protect\cite{ham90}.
\mbox{} }}

The resulting Schr\"odinger equivalent potential (\ref{pot}) is
shown in Fig.~1 as a function of the nucleon kinetic energy with
respect to the nuclear matter at rest in comparison to the data
from Hama et al. \cite{ham90} (open squares). The experimental
increase of the Schr\"odinger equivalent potential up to $E_{kin}
= 1$ GeV is decribed quite well; then the potential decreases and
is set to zero above 3.5 GeV.

For the transition rate in the collision term of the transport
model we employ in-medium cross sections as in Ref. \cite{effe}
that are parameterized in line with the corresponding experimental
data for $\sqrt{s} \leq \sqrt{s_{sw}}$. For higher invariant
collision energies we adopt the Lund string formation and
fragmentation model \cite{lund} as incorporated in the HSD
transport approach \cite{ehe96} which has been used extensively
for the description of particle production in nucleus-nucleus
collisions from SIS to SPS energies \cite{cass99}. In the present
relativistic transport approach (RBUU) as in Ref. \cite{sahu98} we
ex\-pli\-cite\-ly propagate nucleons and $\Delta$'s as well as all
baryon resonances up to a mass of 2 GeV with their isospin degrees
of freedom \cite{effe,teis96}. Furthermore, $\pi, \eta$, $\rho$,
$\omega$, $K, \bar{K}$ and $\sigma$ mesons are propagated, too,
where the $\sigma$ is a short lived effective resonance that
describes s-wave $\pi \pi$ scattering. For more details we refer
the reader to Refs.~\cite{effe,teis96} concerning the low energy
cross sections and to Refs. \cite{ehe96,cass99} with respect to
the implementation of the string dynamics.

\subsection{Transverse mass spectra of protons}
In Fig.~2  we show the dependence of the calculated proton
transverse mass spectra in a central collision of Au + Au at 11.6
$A$ GeV for b $< 3.5$ fm for $\sqrt{s_{sw}}$ = 2.6 GeV (dotted
histograms) and 3.5 GeV (solid histograms) in comparison to the
experimental data of the E802 collaboration \cite{E802}. A cascade
calculation (crosses) is shown additionally for $\sqrt{s_{sw}}$ =
3.5 GeV to demonstrate the effect of the mean-field potentials
which lead to a reduction of the transverse mass spectra  below
0.3 GeV. As expected, the transverse mass spectrum is softer for
smaller $\sqrt{s_{sw}}$ due to the larger number of degrees of
freedom in the string model relative to the resonance model.

\hskip 0.95in \psfig{figure=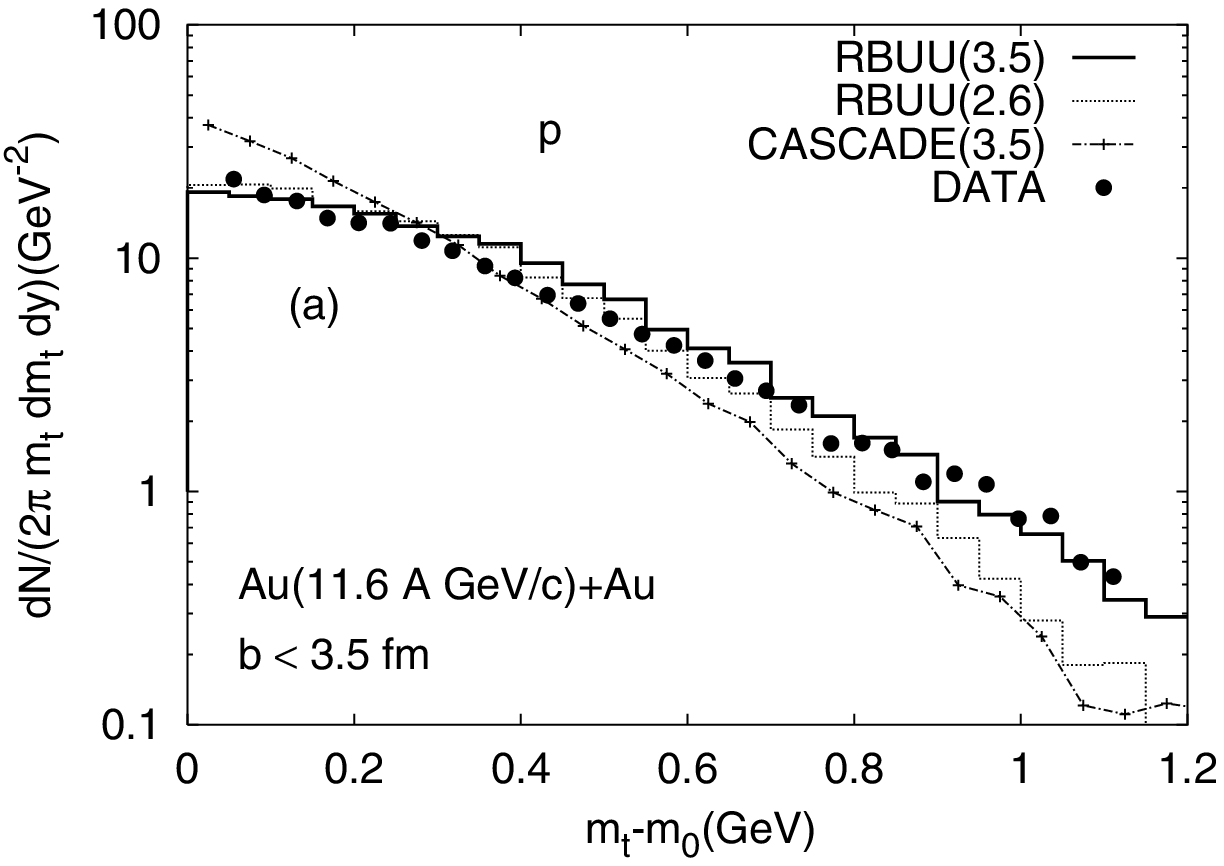,width=6.5cm}\\
{\noindent \small {{\bf Fig. 2:}  The transverse mass
spectra of protons for $Au+Au$ collisions at $b < 3.5$ fm. The
solid line and the dot- dashed line with crosses are results for
$\sqrt{s_{sw}}$=3.5 GeV with and without nuclear potentials,
respectively. The dotted line RBUU(2.6) is for $\sqrt{s_{sw}}$=2.6
GeV. The data points are taken from the E802 collaboration
\cite{E802}. \mbox{} }}

\vskip 0.1in

 We note that strings may be regarded as hadronic
excitations in the continuum of lifetime $t_F \approx$ 0.8 fm/c
(in their rest frame) that take over a significant part of the
incident collision energy by their invariant mass. They decay
dominantly to light baryons and mesons and only to a low extent to
heavy baryon resonances. Thus the  number of particles for fixed
system time is larger for string excitations than for the
resonance model where several hadrons propagate as a single heavy
resonance which might be regarded as a cluster of a nucleon + $n$
pions. As a consequence the translational energies are suppressed
in string excitations and, as a result, the temperature as well as
the pressure are smaller when exciting strings.

From the above comparison with the experimental transverse mass
spectra for protons we find $\sqrt{s_{sw}} \approx$ 3.5 GeV, which
implies that binary final baryon channels should dominate up to
$\sqrt{s} \approx$ 3.5 GeV which corresponds to a proton
laboratory energy of about 4.6 GeV for $pp$ collisions.

\section{SIS energies}

\subsection{Transverse flow}

Within the RBUU model described above we now calculate the
transverse flow for various systems and beam energies and analyse
the dependence on different quantities. The flow $F$ is defined as
the slope of the transverse momentum distribution at midrapidity,
\be
\label{flow}
 F={\frac{d\langle p_x\rangle}{dy}}_{|y=y_0}, \ee
which is essentially generated by the {\it participating} matter
in the 'fireball' \cite{hom98}. The latter finding explains why
flow does not clearly distinguish between an EoS with and without
momentum-dependent forces. Since the fireball contains the stopped
matter, the relative momenta in the fireball (besides the
unordered thermal motion) are small. Only when applying additional
cuts, e.g. on high transverse momenta \cite{Dani,bas97}, i.e. by
selecting particles escaping early from the fireball, or selecting
mainly participant or spectator particles by appropriate
$\Theta_{cm}$-Cuts \cite{Cro97}, a difference between the
momentum-dependent and momentum-independent EoS can be
established.

The FOPI data on proton flow \cite{her96} indicate a decrease of
sidewards flow above 1 A GeV incident energy following the well
known logarithmic increase at low energies. Using standard
potential parameterizations, both nonrelativistic \cite{GBdG87}
and relativistic \cite{sahu98}, this behavior cannot be understood
within conventional transport models. In the latter the optical
potential stays constant or even increases at high momenta and
therefore the repulsion generated from the momentum-dependent
forces in a HIC gives rise to a significant contribution to the
flow signal. However, since the nucleon-nucleus optical potential
is only known up to 1 GeV experimentally \cite{ham90}, it was
proposed that this decrease of flow above 1 A GeV might indicate a
decrease of the optical potential at high relative momenta or at
high baryon density \cite{sahu98}.

In Ref. \cite{BKL91} the transverse momentum of the baryons has
been disentangled into a collisional part, a mean-field part and a
part originating from the Fermi-motion of the particles,
\be
p_t = p_t^{\rm coll} + p_t^{\rm MF} + p_t^{\rm Fermi} \;, \ee
where $p_t^{\rm Fermi}$ practically does not contribute at
midrapidity. Thus disentangling the flow signal into a collisional
and a potential part, it turns out that $\sim$50 \% of the flow
stems from the particle - particle collisions while roughly
another 50 \% are generated by the potential repulsion at 1 A GeV.
Fig. 3 shows these two contributions for flow for the system
$Ni+Ni$ at b=4 fm in comparison to the experimental data using
different EoS denoted by 'hard', 'medium' and 'soft'. Both the
collisional ''background'' and the potential part rise up to 1 A
GeV incident energy and remain constant above, whereas the data
indicate a decrease above 1 A GeV. As seen from Fig. 3 the $Ni+Ni$
data are described reasonably by both a 'soft' and 'medium' EoS,
while a cascade calculation fails substantially.

\vskip 0.1in \hskip 0.65in \psfig{figure=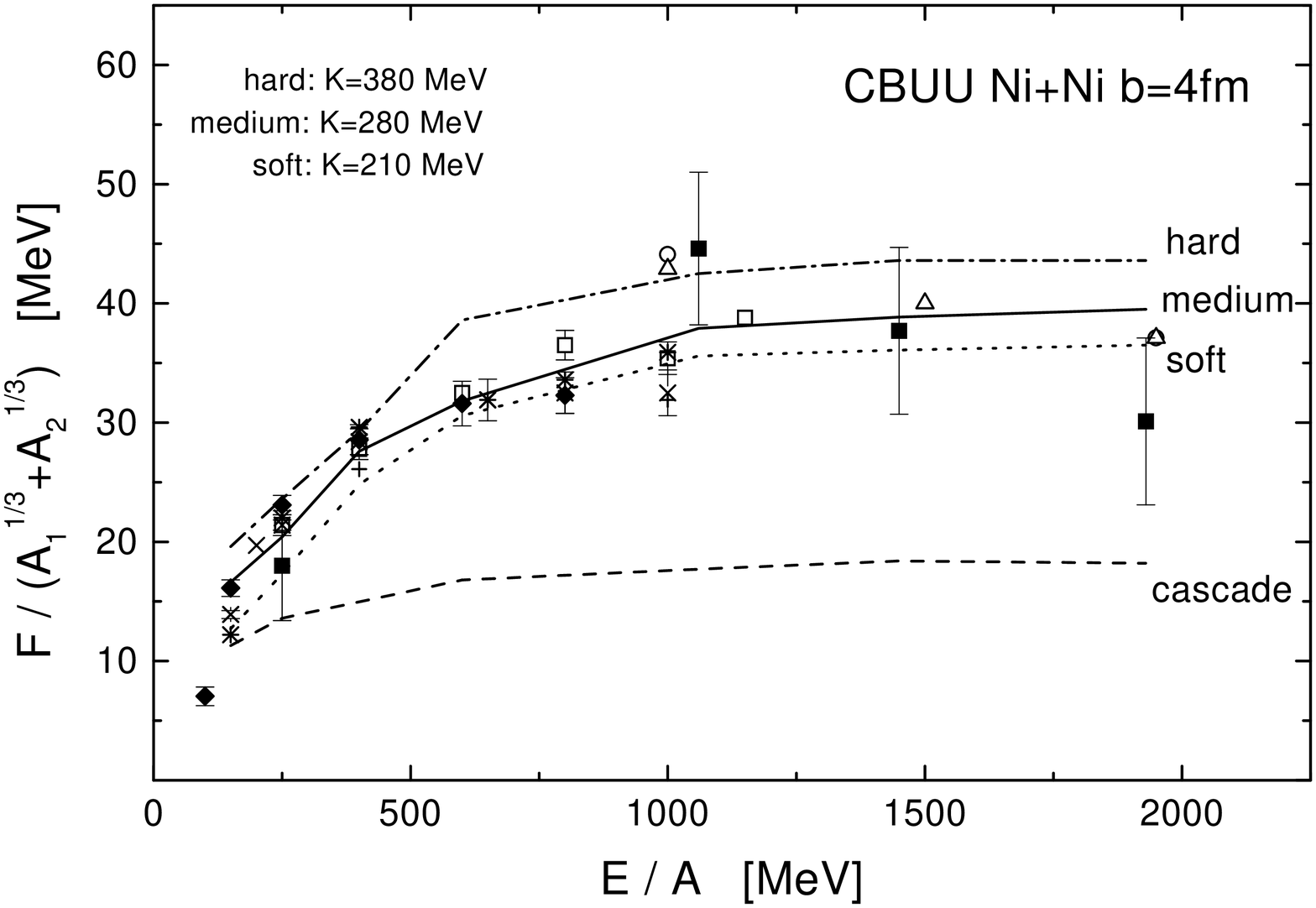,width=8cm}
\vskip 0.05in {\noindent \small {{\bf Fig. 3:} Transverse flow for
a $Ni+Ni$ collision at b=4 fm as calculated within the RBUU model
in cascade mode (dashed lower line) and for equations of state
with different compressibilities $K$ in comparison to data from
EOS, Plastic Ball and FOPI as compiled by Ref.~\cite{her96}.
\mbox{} }} \label{trfl_eos} \vskip 0.1in

A note of caution has to be added here: The flow $F$ (\ref{flow})
not only depends on the baryon self-energies $U_S$ and $U_\mu$ but
also on the number of (resonance) degrees of freedom above about 1
A GeV as first pointed out by Hombach et al. \cite{hom98}. This
observation will become crucial at AGS energies.

\subsection{Radial flow}

Radial flow has been discovered \cite{Jeong94}
when analyzing the flow pattern of very central events of RHIC.
In contrast to transverse flow up to about 70 \% of the
incident energy (stored in the hot compressed fireball)
is released as ordered radial expansion of the
nuclear matter. Thus the hope is to extract information
especially on the compressibility of the EoS via the magnitude of the
radial flow.

Experimentally the radial flow is characterized or fitted in terms of
the Siemens-Rasmussen formula
\cite{SR79}
\be
\label{SR} \frac{d^3N}{dEd^2\Omega} \sim p \cdot e^{-\gamma E/T}
\left\{\frac{\sinh \alpha}{\alpha} \cdot (\gamma E + T) - T \cdot
\cosh\alpha \right\} \ee with $\gamma=(1-\beta^2)^{-1/2}$ and
$\alpha=\gamma\beta p/T$, while $\beta$ denotes the flow velocity
and $T$ characterizes some temperature. We follow the same
strategy in our RBUU calculations and apply a least square fit to
the RBUU nucleon spectra using Eq.~(\ref{SR}).

\vskip 0.15in \psfig{figure=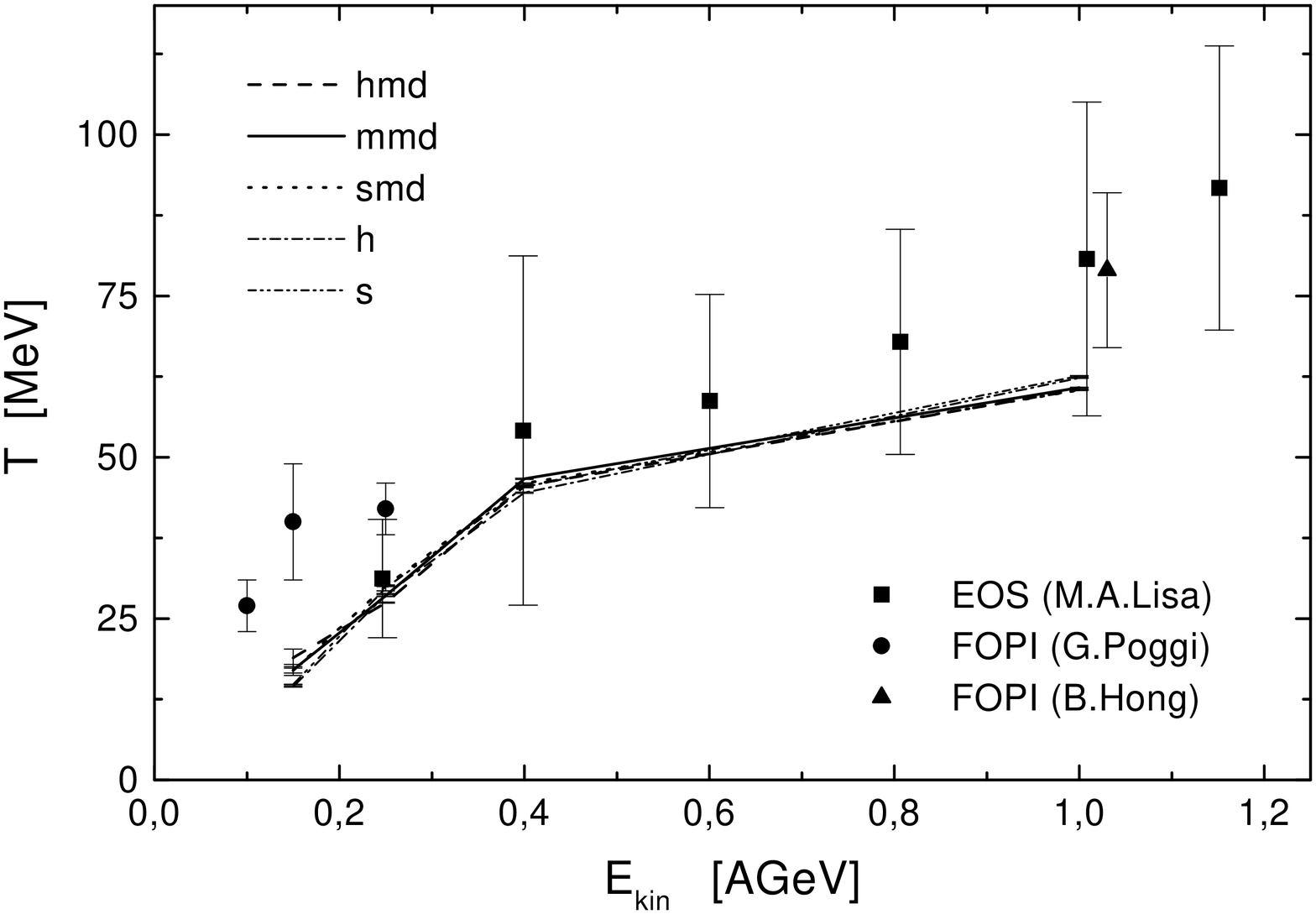,width=5.8cm} \hskip 0.1in
\psfig{figure=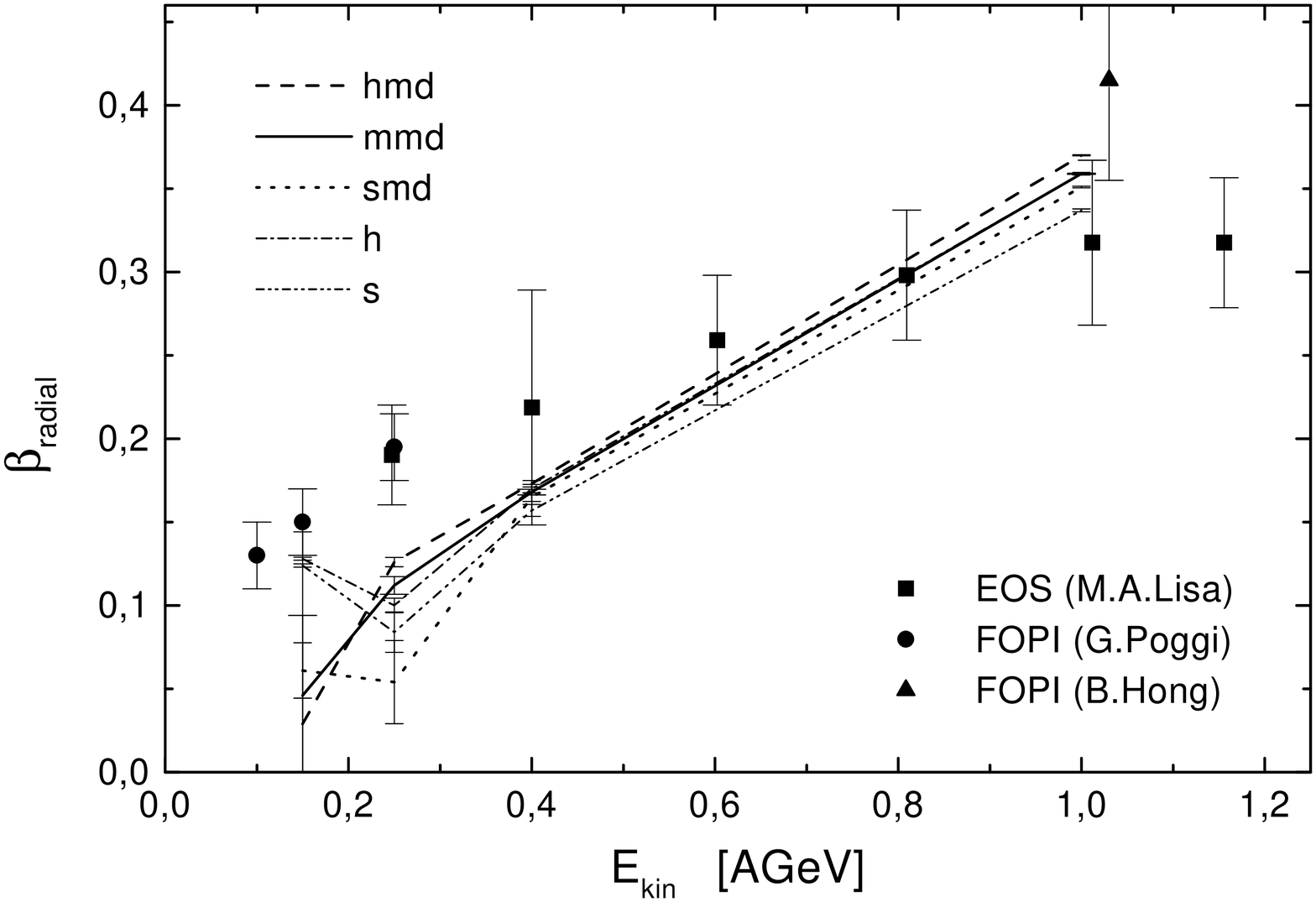,width=5.8cm}
 \vskip 0.15in {\noindent \small
{{\bf Fig. 4:} The temperature (l.h.s.) and radial flow velocity
(r.h.s.) for central $Au+Au$ collisions evaluated via
Eq.~(\ref{SR}) from the RBUU calculations in comparison to the
experimental data from Refs.~\cite{Lisa95,Poggi95,Hong97}. The
symbol 's' denotes a soft EoS without momentum dependent forces,
'h' a hard EoS and 'smd', 'mmd' and 'hmd' correspond to a soft,
medium and hard momentum dependent EoS, respectively. \mbox{} }}
\label{rad} \vskip 0.1in

 The results of the RBUU calculations for central
$Au+Au$ collisions are shown in Fig. 4 as a function of the
bombarding energy in comparison to the data from
\cite{Lisa95,Poggi95,Hong97}. We find that the 'temperature' $T$
is systematically underpredicted in all schemes investigated (soft
and hard EoS, with and without momentum dependent forces), and
that the flow velocities are not correctly reproduced, being too
low at low energy and crossing the experimental data around 800 A
MeV. This might indicate a strong binding from the potential which
gives not enough repulsion at high densities and overcompensates
the collisional pressure from the fireball. However, the nucleon
spectra resulting from the RBUU calculation show a strong
non-thermal component at low incident energies and are thus in
contradiction to the physical picture behind Eq.~(\ref{SR}) which
assumes an isentropic expansion of a thermal equilibrated source.
In Ref. \cite{hom98_1} the degree of equilibration in a HIC has
been investigated as a function of the incident energy and the
system mass and it has been found that even the most massive
systems like $Au+Au$ do not equilibrate at low energies.

\section{AGS energies}

\subsection{Sidewards flow}
We now turn to the AGS energy regime from 1 -- 11 AGeV. The
calculations are performed for the impact parameter $b=6 fm$ for
$Au+Au$ systems, since for this impact parameter we get the
maximum flow which corresponds to the multiplicity bins $M3$ and
$M4$ as defined by the Plastic Ball collaboration \cite{dos87} at
BEVALAC/SIS energies. In Fig.~5 (l.h.s.) the transverse flow
(\ref{flow}) is displayed in comparison to the data from
Refs.~\cite{her96,rei97,EoS}\ for $Au + Au$ systems. The solid
line (RBUU with $\sqrt{s_{sw}}$ = 3.5 GeV) is obtained with the
scalar and vector self energies as discussed above, Eq.
(\ref{form}). The dotted line (CASCADE with $\sqrt{s_{sw}}$ = 3.5
GeV) corresponds to cascade calculations for reference in order to
show the effect of the mean field relative to that from
collisions. We observe that the solid line (RBUU, cf. Fig.~1) is
in good agreement with the flow data \cite{EoS} at all energies;
above bombarding energies of 6 A GeV the results are practically
indentical to the cascade calculations showing the potential
effects to cancel out.

We note that the sideward flow shows a maximum around 2 A GeV for
$Au+Au$ and decreases continuously at higher beam energy ($\ge$ 2
A GeV) without showing any explicit minimum as in Ref.
\cite{Rischke}. This is due to the fact that the repulsive force
caused by the vector mean field  decreases at high beam energies
(cf. Fig. 1) such that in the initial phase of the collision there
are no longer strong gradients of the potential within the
reaction plane. In subsequent collisions, which are important for
$Au+Au$  due to the system size, the kinetic energy of the
particles relative to the local rest frame is then in a range
($E_{kin} \leq$ 1 GeV) where the Schr\"odinger equivalent
potential (at density $\rho_0$) is determined by the experimental
data~\cite{ham90}. We thus conclude that for the sideward flow
data up to  11 A GeV one needs a considerably strong vector
potential at low energy and that one has to reduce the vector mean
field at high beam energy in line with Fig. 1. In other words,
there is only a weak repulsive force at high relative momenta or
high densities.

 \psfig{figure=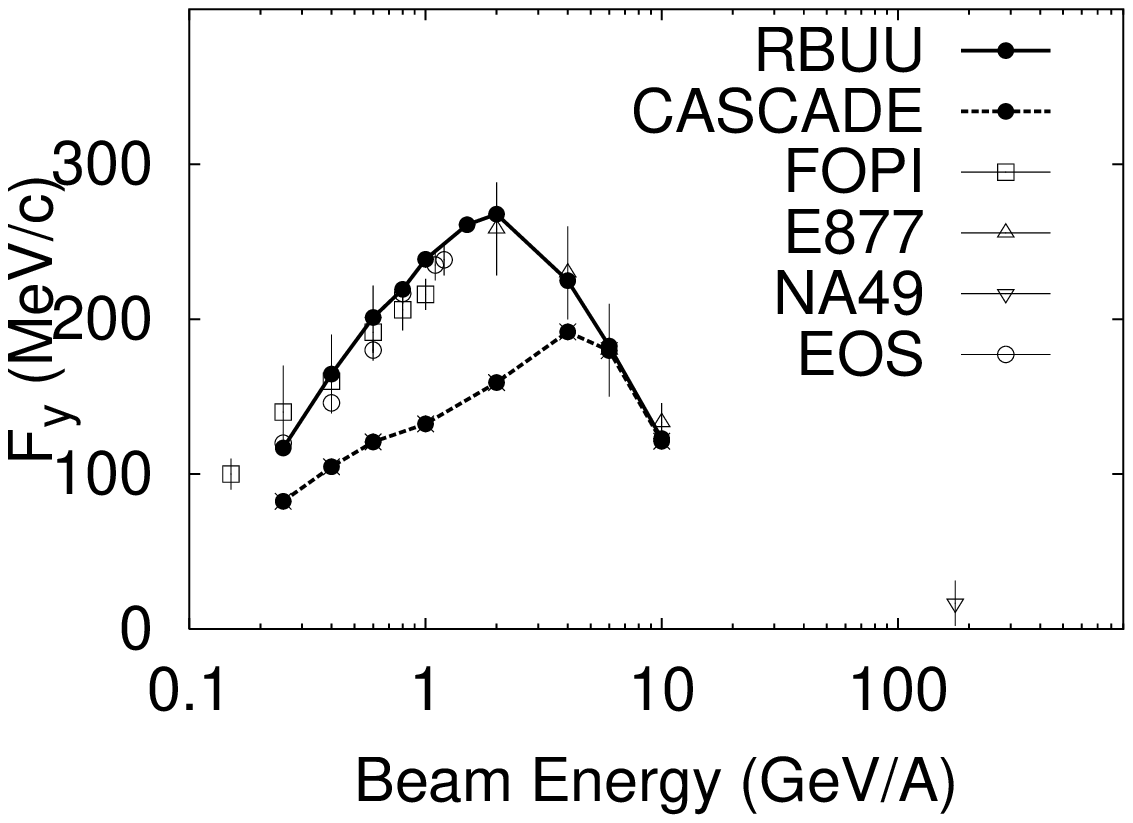,width=6cm} \hskip 0.2in
 \psfig{figure=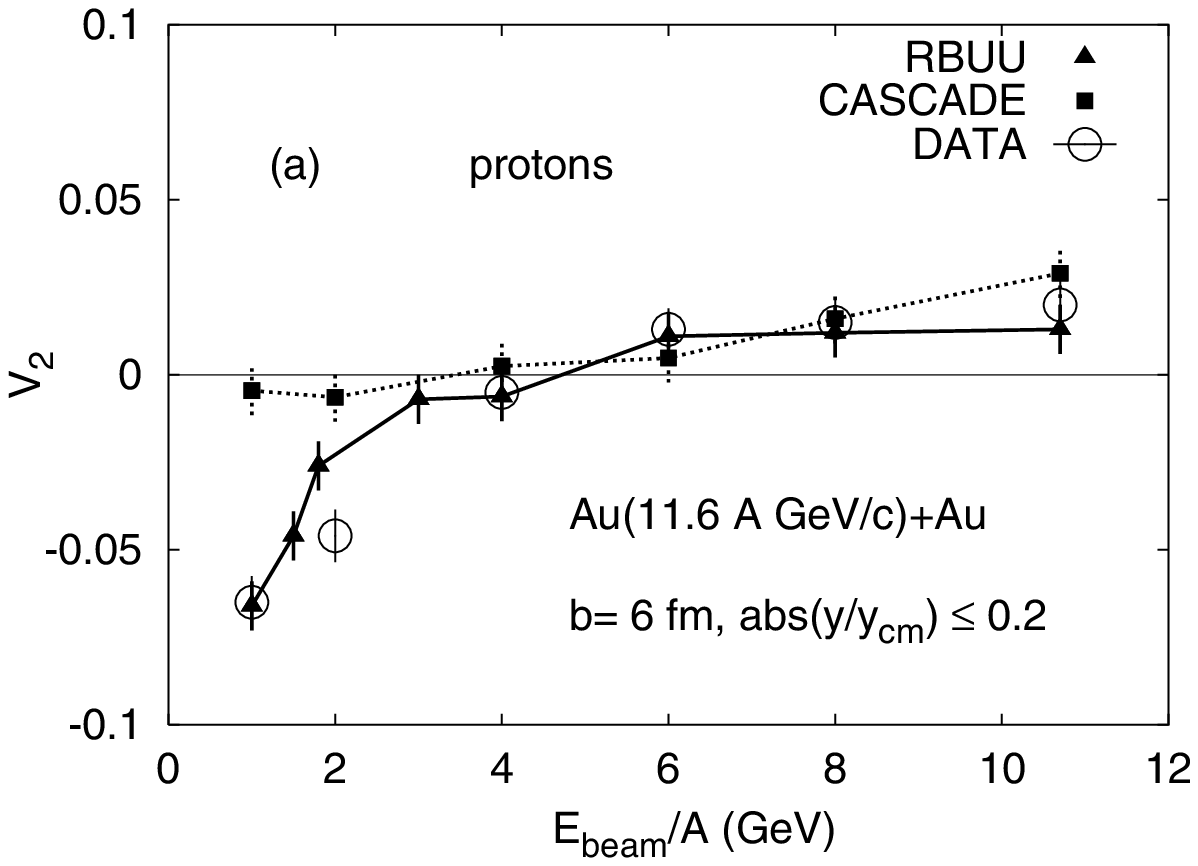,width=5.5cm} \vskip 0.15in
{\noindent \small {{\bf Fig. 5:} (l.h.s.) The sideward flow $F(y)$
as a function of the beam energy per nucleon for $Au+Au$
collisions at $b=6$ fm from the RBUU calculations. The solid line
results for the parameter set RBUU, the dotted line for a cascade
calculation with $\sqrt{s_{sw}}$ = 3.5 GeV. The data points are
from the FOPI and EoS Collaborations \protect\cite{her96,EoS}.
(r.h.s.) The elliptic flow $v_2$ of protons versus the beam energy
per nucleon for $Au+Au$ collisions at $b=6$ fm from the RBUU
calculations. The solid line results for the parameter set RBUU,
the dotted line for a cascade calculation with $\sqrt{s_{sw}}$ =
3.5 GeV. The data points are from the EoS Collaboration
\cite{E877,E895}. \mbox{} }} \vskip 0.1in

Another aspect of the decreasing sideward flow can be related to
the dynamical change in the resonance/string degrees of freedom as
already discussed above. For instance, for $\sqrt{s_{sw}} = 2.6$
GeV the calculated flow turns out to be smaller than the data
above 1.5 A GeV and approaches the cascade limit already for
$\approx$ 3-4 A GeV. This is due to the fact that in strings the
incident energy is stored to a larger extent in their masses and
the translational energy is reduced accordingly.

\subsection{Elliptic flow}
Apart from the in-plane flow of protons the out-of-plane
collective flow provides additional information and constraints on
the nuclear potentials involved. In this respect the elliptic flow
for protons
\begin{equation}
v_2 = \langle (P_x^2-P_y^2)/(P_x^2+P_y^2) \rangle
\end{equation}
 for $|y/y_{cm}| \leq 0.2 $ is shown in Fig. 5 (r.h.s.) as
a function of incident energy for $Au + Au$ collisions at $b$ = 6
fm. The solid line (RBUU with $\sqrt{s_{sw}}$ = 3.5 GeV) is
obtained with the same mean fields as discussed before while the
dotted line (CASCADE with $\sqrt{s_{sw}}$ = 3.5 GeV) stands again
for the cascade results. The flow parameter $v_2$ changes its sign
from negative at low energies ($\le 5 A$ GeV) to positive elliptic
flow at high energies ($\ge 5 A$ GeV).

This can be understood as follows: At low energies the squeeze-out
of nuclear matter leads to a negative elliptic flow since
projectile and target spectators distort the collective expansion
of the 'fireball' in the reaction plane. At high energies the
projectile and target spectators do not hinder anymore the
in-plane expansion of the 'fireball' due to their high velocity
($\approx c$); the elliptic flow then is positive. The competition
between squeeze-out and in-plane elliptic flow at AGS energies
depends on the nature of the nuclear force as pointed out already
by Danielewicz et al. \cite{dani98}. We note, however, that in our
calculation with the momentum-dependent potential (Fig. 1) we can
describe both the sideward as well as elliptic flow data
\cite{E877,E895} simultaneously without incorporating any phase
transition in the EOS as in Ref. \cite{dani98}.

In the cascade calculation the elliptic flow from squeeze-out is
weaker due to the lack of a nuclear force which demonstrates the
relative role of the momentum-dependent nuclear forces on the
$v_2$ observable below bombarding energies of about 5 A GeV.

\section{Summary}
In this contribution we have explored the dependence of transverse
and radial flow signals on various model inputs - that are related
to the nuclear EoS - using the coupled channel RBUU model. We find
that the mass distribution of the resonances included in the model
plays an important role for the description of transverse flow
above \mbox{1 A GeV}. For the radial flow we have concentrated on
the difference between the results for the flow temperature $T$
and flow velocity $\beta$ when using different EoS. However, no
sizeable sensitivity to the compressibility of the EoS could be
established.

On the other hand, we found that in order to reach a consistent
understanding of the nucleon optical potential up to 1 GeV, the
transverse mass distributions of protons at AGS energies as well
as the excitation function of sidewards and elliptic flow
\cite{E877,EoS} up to 11 A GeV, the strength of the vector
potential has to be reduced in the RBUU model at high relative
momenta and/or densities. Otherwise, too much flow is generated in
the early stages of the reaction and cannot be reduced at later
phases where the Schr\"odinger equivalent potential is
experimentally known. This constrains the parameterizations of the
explicit momentum dependence of the vector and scalar mean fields
$U_\mu$ and $U_S$ at high relative momenta.

In addition, we have shown the relative role of resonance and
string degrees of freedom at AGS energies. By reducing the number
of degrees of freedom via high mass resonances one can build up a
higher pressure and/or temperature of the 'fireball' which shows
up in the transverse mass spectra of protons as well as in the
sidewards flow \cite{sahu99}. A possible transition from resonance
to string degrees of freedom is indicated by the RBUU calculations
at invariant baryon-baryon collision energies of $\sqrt{s}
\approx$ 3.5 GeV which corresponds to a proton laboratory energy
of about 4.6 GeV. Due to Fermi motion of the nucleons in $Au + Au$
collisions the transition from resonance to string degrees of
freedom becomes smooth and starts from about 3 A GeV; at 11 A GeV
practically all initial baryon-baryon collisions end up in
strings, i.e. hadronic excitations in the continuum that decay to
hadrons on a time scale of about 0.8 fm/c in their rest frame.
This initial high-density {\it string matter} (up to 10 $\rho_0$
at 11 $A$ GeV) should not be interpreted as {\it hadronic matter}
since it implies roughly 5 constituent quarks per fm$^3$, which is
more than the average quark density in a nucleon.

It is interesting to note that at roughly 4 A GeV in central $Au +
Au$ collisions the ratio $K^+/\pi^+$ is enhanced experimentally
relative to transport calculations \cite{cass98}. One might
speculate that a restoration of chiral symmetry could be
responsible for the softer collective response as well as enhanced
strangeness fraction.

\end{document}